\documentclass[%
reprint,
amsmath,amssymb,
aps,
floatfix,
]{revtex4-2}

\usepackage{xcolor}
\usepackage{graphicx}
\usepackage{dcolumn}
\usepackage{bm}
\usepackage{hyperref}

\usepackage{amsmath}
\usepackage{float}
\usepackage{amsfonts}
\usepackage{soul}
\usepackage[normalem]{ulem}
\usepackage{placeins}

\hypersetup{
	colorlinks=true,
	linkcolor=blue,
	filecolor=black,
	anchorcolor=black,
	citecolor=blue,
	filecolor=black,
	urlcolor=blue
}

\newcommand{\im}{{\rm i}}

\begin{document}
	
	\preprint{APS/123-QED}
	
	\title{Attosecond Light Skyrmion Pulses via High Harmonic Generation} \author{David Marco\textsuperscript{1,2,3}}	\email{davidmarco@usal.es}
	\author{Luis Plaja\textsuperscript{1,2}} \author{Carlos Hernández-García\textsuperscript{1,2}}

	\affiliation{%
		\textsuperscript{1}Grupo de Investigación en Aplicaciones del Láser y Fotónica, Departamento de Física Aplicada, Universidad de Salamanca, E-37008 Salamanca,
	Spain\\
	\textsuperscript{2}Unidad de Excelencia en Luz y Materia Estructuradas (LUMES), Universidad de Salamanca, Salamanca, Spain\\
    	\textsuperscript{3}Instituto de Bioingeniería, Universidad Miguel Hernández de Elche, 03202 Elche, Spain\\
	}%
	
	\begin{abstract}
Paraxial light skyrmions are topological configurations that map a spatial domain of the field onto the full Poincaré sphere of polarization states. While optical skyrmions have been explored in continuous-wave regimes, their realization in the ultrafast domain remains open. Here we demonstrate that attosecond skyrmion pulses can be generated via high-harmonic generation. Advanced simulations combining single-atom strong-field theory and macroscopic propagation reveal that an infrared linearly polarized vector beam with fractional orbital angular momentum produces extreme-ultraviolet harmonic fields with nearly identical skyrmion polarization distributions across a broad spectral range. Using 1.2 $\mu$m driving fields and experimentally feasible spectral filtering, we show that the coherent superposition of consecutive harmonics centered at 70 eV yields a train of skyrmion pulses with $\sim500$ attoseconds duration. Our results open opportunities to use structured attosecond light with topological polarization textures in fields as ultrafast control, imaging and spectroscopy.
	\end{abstract}
	
	\maketitle

\begin{figure*}
\centering
\includegraphics[width=0.98\textwidth]{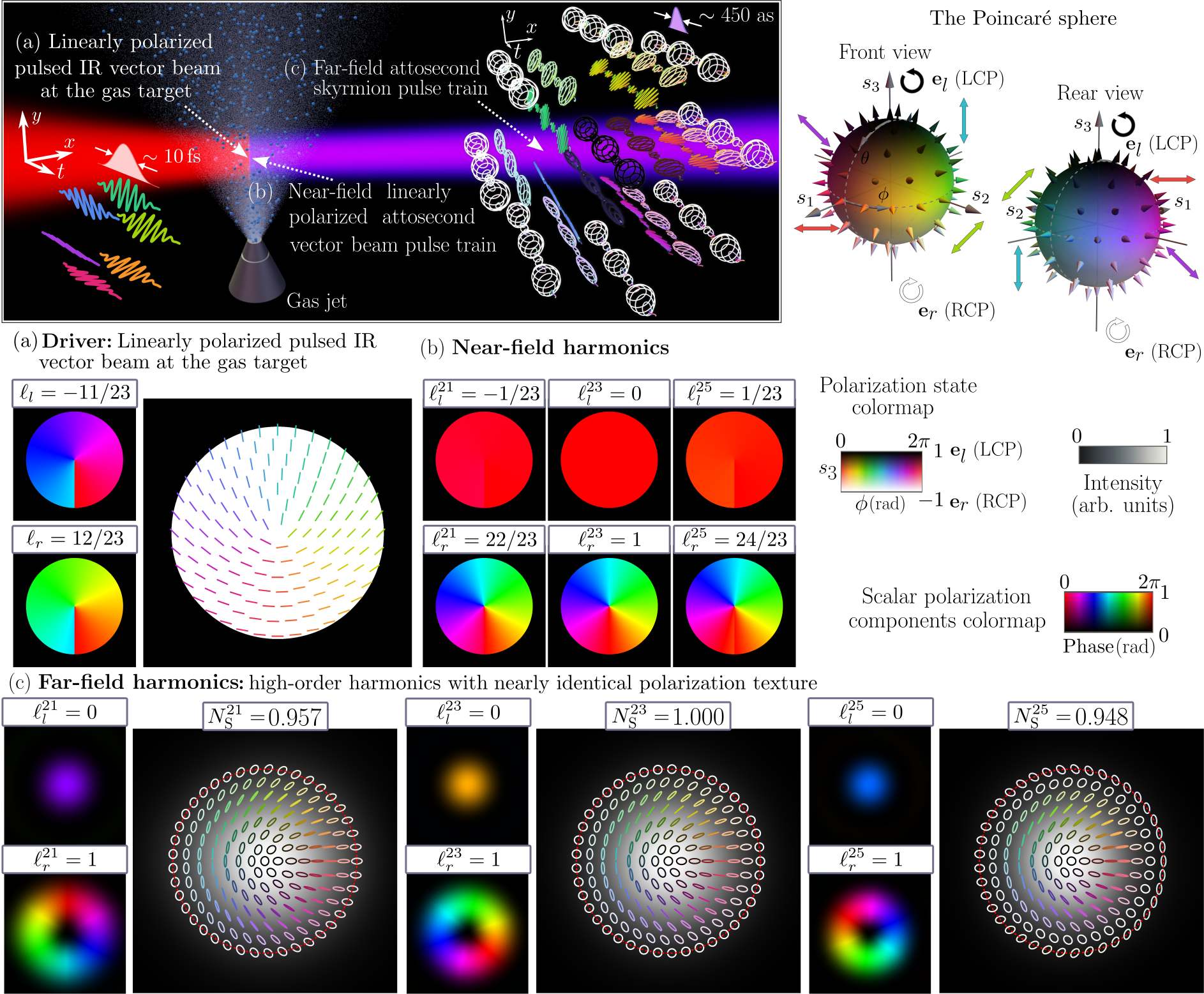}
\caption{Generation of an attosecond EUV skyrmion pulse trains from an IR linearly polarized fs vector beam via HHG. The instantaneous polarization state of the pulse train is encoded by a color map that assigns each point to its location on the Poincaré sphere.
(a) Input driver at the gas target: intensity–phase maps of the LCP and RCP components, together with the intensity–polarization map designed to produce a skyrmion in the 23rd harmonic. 
(b) Near-field intensity–phase maps of the LCP and RCP components for harmonics $q=21,23,25$, where the polarization distribution equals that of the driver. 
(c) Far-field intensity–phase maps of the circular components for harmonics $q=21,23,25$, along with the corresponding intensity–polarization maps. All three harmonics evolve into nearly equal skyrmion structures with $N^{q}_{\mathrm{S}}\approx1$ within a common closed contour of 0.34 mrad in the far field, indicated by the red dashed line. 
}
\label{fig:figure1}
\end{figure*}

Skyrmions are field configurations with nontrivial topology, where a spatial region is mapped onto an entire spherical parameter space \cite{Skyrme, prop_invariant_meron_lattices}. Their topological properties are quantified by the Skyrme number $N_\mathrm{S}$, whose magnitude gives the number of sphere coverings.
Skyrmionic textures have been identified in systems like magnetic media \cite{magnetic_skyrmions_review}, superfluids \cite{meron_lattice_superfluid_book}, Bose–Einstein condensates \cite{hansen2016singular}, acoustic \cite{Ge_Acoustic,Muelas} and water waves \cite{water_skyrmions}, and in various aspects of light polarization \cite{review_optical_skyrmions_Shen,Dennis_hopfion,first_optical_skyrmions,skyrmion_spin_evanescent,skyrmions_Rodrigo_Pisanty,optical__merons_PRL,mata2025skyrmionic,marco20254d}. In particular, skyrmions arise in paraxial light fields containing all possible elliptical polarization trajectories in the transverse profile  \cite{full_Poincare_beams,Poincare_skyrmions,cartography_skyrmions,prop_invariant_meron_lattices}. These trajectories are fully represented as points on the Poincaré sphere of polarization states, where the poles correspond to the two orthogonal circular states and the equator to linear states (Fig. ~\ref{fig:figure1}). Points in physical space mapping to circular polarization states are known as C-points.
Unlike phase singularities in scalar optical vortices, C-points can be bright as they represent singularities in the orientation of the polarization ellipse in the circle limit, and have found recent applications in sensing \cite{zhou2024controllable}. Paraxial skyrmions are being also applied in  complex media \cite{wang2024topological}, quantum optics \cite{ornelas2024topological}, topological materials \cite{mkhumbuza2025topological, mitra2025topological} or imaging \cite{peters2025seeing}.

Up to now, optical skyrmions have been generated in the continuous-wave or long-pulse regimes, while their generation and applications in the ultrafast domain remain largely unexplored. On the other side, although the production of light pulses down to the attosecond timescale is well established since the early 2000's \cite{Paul2001, Hentschel2001}, the generation of structured fields at such timescales has only been explored during the last decade. The key challenge is how to structure light in the high-frequency regime, where most optical techniques become highly inefficient. Recently, x-ray paraxial skyrmions have been generated using free electron lasers, without relying in optical elements \cite{morgan2025poincare}. However, synthesizing attosecond skyrmions requires the production of a high-frequency comb of phase-locked radiation with, ideally, identical skyrmion characteristics. In this context, the nonlinear process of high harmonic generation (HHG) \cite{McPherson1987, Ferray1988} has already been proven to be a robust method for generating  simpler forms of attosecond structured light pulses. In HHG an intense infrared (IR) femtosecond (fs) driver is up-converted into a \emph{plateau} of harmonics with similar intensity and regular phase relationship, which allows their synthesis into attosecond pulses \cite{Paul2001, Hentschel2001}. Through HHG, some particular forms of structured IR fields can be up-converted to structured extreme-ultraviolet (EUV) beams. As a result, EUV attosecond structures such as vortices \cite{Hernandez-Garcia2013, Geneaux2016, Dorney2019, delasHeras2024}, vector beams \cite{Hernandez-Garcia_17_vectorbeamsHHG}, self-torque beams \cite{Rego2019}, and spatiotemporal optical vortices \cite{Martin-Hernandez2025a, Martin-Hernandez2025b} have been successfully generated. Thus, HHG stands as an excellent potential candidate for producing far more complex forms of ultrafast structured fields, as attosecond skyrmions.

However, the up-conversion of IR paraxial skyrmions though HHG is far from trivial, as the harmonic conversion efficiency depends critically on the polarization state, decreasing substantially with the driver ellipticity \cite{Budil1993}. 
As such, the up-conversion of light structures containing polarization states other than linear is challenging. Therefore, an IR paraxial skyrmion does not directly up-convert into an EUV harmonic skyrmion. Nonetheless, by properly modifying the IR driving field properties, high-order harmonics with elliptical and circular polarization states have been produced \cite{Fleischer2014, Hickstein2015, Huang2018, Brooks2025}. Indeed, two-color schemes have been proposed to generate EUV skyrmions with harmonic-dependent polarization distributions and distinct $N_\mathrm{S}$ \cite{fang2023generation}. Yet, producing attosecond skyrmion pulses further requires a spectral comb where the harmonics share the same polarization distribution while maintaining a regular phase relationship.

In this Letter, we demonstrate that attosecond skyrmion pulses can be obtained in HHG by properly structuring a linearly polarized IR vector driving beam. Our simulations demonstrate that the use of two counter-rotating beams with properly chosen fractional orbital angular momentum (OAM) drives far-field harmonics that, within a common fixed circular contour, exhibit polarization textures with nearly identical Skyrme number. The coherent superposition of the harmonic comb yields an attosecond pulse train with a temporally nearly constant skyrmion polarization texture (Fig.~\ref{fig:figure1}). 
In particular, as a potential experimental implementation, we propose that, driving HHG with a 1.2 $\mu$m driving and using of a combination of Zr and Al filters, skyrmion pulses of $\sim$500 attoseconds and a central photon energy of $\sim$70 eV can be readily produced.

The proposed HHG configuration is depicted in Fig.~\ref{fig:figure1} (top). HHG is driven by two counter-rotating IR beams with fractional OAM, that results in a linearly polarized vector beam at the gas jet, given by 
\begin{equation}
\mathbf{A}(\rho,\varphi)
= A(\rho) \left[e^{\im \ell_l \varphi} \, \mathbf{e}_l + e^{\im \ell_r \varphi} \, \mathbf{e}_r \right]=A(\rho) e^{\im \ell_+ \varphi} \mathbf{p}(\varphi),
\label{eq:driving_field}
\end{equation}
where $\mathbf{e}_l,\mathbf{e}_r=(1\pm \im)/\sqrt2$ are the basis vectors of the left- and right-circularly polarized (LCP and RCP) components carrying topological charges of $\ell_{l,r}$, $\ell_\pm=(\ell_l\pm\ell_r)/2$, and $\mathbf{p}(\varphi)= e^{\im \ell_- \varphi} \, \mathbf{e}_l + e^{-\im \ell_-\varphi} \, \mathbf{e}_r$ denotes a spatially variant linear polarization state, whose local orientation angle is defined by $\ell_-$. We consider a driving amplitude profile, $A(\rho)$, corresponding to a circular aperture with uniform amplitude ($A=1$ for $\rho<\rho_0$ and $A=0$ otherwise, with $\rho_0$ denoting the aperture radius). Note that to preserve the linearly polarized state that optimizes the HHG efficiency, the two circular components must overlap spatially.

 To predict the conditions for  generating skyrmionic harmonic polarization textures, we consider the thin slab model (TSM) for HHG \cite{Hernandez-Garcia_2015}, which treats the gas target as an infinitesimally thin layer of emitters. The TSM models the near-field $q$-th-order harmonic in terms of the driving field---the harmonic amplitude scales non-perturbatively with that of the driving with power of $p<q$, whereas its phase is multiplied by $q$ \cite{l1992calculations}. 
 The model includes an additional intrinsic dipole phase term arising from the electronic excursion. This phase, being proportional to the local driving intensity, can be neglected for the uniform circular aperture considered. 
 When applied to Eq.~\eqref{eq:driving_field}, the near-field q-th order harmonic profile is then given by
%
\begin{align}
\mathbf{A}^{q}(\rho,\varphi) 
&= A^p(\rho) \left[ e^{\im \ell^q_{l}\varphi} \, \mathbf{e}_l + e^{\im \ell^q_{r}\varphi} \, \mathbf{e}_r \right], \\
\ell^q_{l,r}
&= \frac{\ell_l(q \pm 1) + \ell_r(q \mp 1)}{2}.
\label{eq:output_charges}
\end{align}
Reversing Eq.~\eqref{eq:output_charges}, we can infer the driving beam topological charges, $\ell_{l,r}$, required to obtain a particular set of harmonic topological charges $\ell^q_{l,r}$, as
\begin{equation}
    \ell_{l,r}= \frac{(1 \pm q)\ell^q_{l} + (1 \mp q)\ell^q_{r}}{2q}.
    \label{eq:input_charges}
\end{equation}
An optical skyrmion with $N_{\mathrm{S}}=1$ can be constructed from LCP and RCP components carrying topological charges 0 and 1, respectively \cite{full_Poincare_beams}. According to Eq.~\eqref{eq:input_charges}, a $q$-th-order harmonic with topological charges $\ell^{q}_{l,r}=0,1$ is obtained by driving HHG with a linearly polarized vector beam carrying $\ell_{l,r}=1\mp q/(2q)$.
Considering the $q=23$ harmonic---that belongs to the plateau if driven in argon---, the required driver charges are $\ell_{l,r}=-11/23,12/23$ (Fig.~\ref{fig:figure1}(a)).
 We note that fractional OAM beams \cite{Ballantine2016}, have been previously used in different scenarios of HHG \cite{Turpin2017, delasHeras2022}. 
Figure~\ref{fig:figure1}(b) illustrates the resulting near-field circular components of the high-order harmonics predicted by the TSM, which share the same intensity distribution as they inherit the linear polarization of the driving beam. Whereas in the near field the high-order harmonics remain linearly polarized, the skyrmion texture arises after propagation of these near-field profiles to a far-field detector.

To confirm the predictions given by the TSM, we show in Fig.~\ref{fig:figure1}(c) the results obtained using our full numerical simulations of HHG in an argon gas jet. Our method combines the quantum description at the single-atom level using the strong-field approximation, and the macroscopic phase-matching of the harmonics through the electromagnetic field propagator \cite{Hernandez-Garcia2010}. This approach has been successfully employed in the theoretical back-up of experiments of HHG driven by structured  beams---see for example \cite{Hernandez-Garcia_17_vectorbeamsHHG, Dorney2019, Rego2019, delasHeras2022, delasHeras2024, Martin-Hernandez2025a}. 
The input IR driving field exhibits the polarization, amplitude and phase distributions presented in Fig.~\ref{fig:figure1}(a), with a beam radius of 42 $\mu$m, and a $\sin$-squared temporal envelope of 7.7 fs full width at half maximum in intensity (FWHM), with a peak intensity of 9.81 $\times$ 10$^{13}$ W/cm$^2$ and central wavelength of 0.8 $\mu$m. 

The far-field polarization distributions, together with the amplitude and phase of the circular components of the harmonics $q=21,23,25$, are shown in Fig.~\ref{fig:figure1}(c). For $q=23$, the LCP (RCP) component presents $\ell^{23}_l=0$ ($\ell^{23}_r=1$). Both components exhibit distinct radially varying intensity distributions, yielding a skyrmion with $N^{23}_\mathrm{S}=1$. For adjacent harmonics, the harmonic-order dependence in the selection rule given by Eq.~\eqref{eq:input_charges} leads to small deviations in the topological charges of their polarization components with respect to the central values $\ell^{23}_{l,r}=0,1$, resulting in slightly noninteger charges (Fig.~\ref{fig:figure1}(b)). Interestingly, after propagation to the far field, these noninteger charges evolve into beams with the same integer topological charges \cite{berry2004optical}, $\ell^{21}_{l,r}=\ell^{25}_{l,r}=0,1$, yielding nearly identical skyrmion polarization distributions. 
The far-field polarization distribution of the adjacent harmonics, however, undergoes a slight linear displacement along the $x$ axis, opposite in direction for harmonics above or below $q=23$. 

The near-field circular aperture---$A(\rho)$ in Eq.~\eqref{eq:driving_field}---induces fast polarization oscillations beyond the harmonic skyrmion boundary, but only in regions of negligible intensity---not represented in Fig.~\ref{fig:figure1}(c). Although the skyrmionic structure does not extend to infinity, a skyrmion for the central harmonic $q=23$ can be obtained by spatially cropping the field to the closed contour, $C$, of the RCP component. We calculate the resulting Skyrme number as
\begin{align}
N_\mathrm{S} &= \frac{1}{4\pi} \iint_C \mathbf{s} \cdot \left[\partial_x \mathbf{s} \times \partial_y \mathbf{s} \right] \; \mathrm{d}x \, \mathrm{d}y,
\end{align}
with
\begin{align}
	\mathbf{s}&=\frac{[
		|A^q_x|^2-|A^q_y|^2, 2\mathrm{Re}(A^{q,*}_x A^q_y),2\mathrm{Im}(A^{q,*}_x A^q_y)]}{|A^q_x|^2+|A^q_y|^2},
\label{eq:stokes}
\end{align}
which denotes the spatially varying normalized Stokes vector across a transverse plane, whose components $\mathbf{s}=[s_1,s_2,s_3]$ are the Cartesian coordinates of a point on the Poincaré sphere representing a specific polarization state, and $A^q_{x,y}$ are the spatially varying Cartesian components of the complex harmonic field.  
Using a fixed closed contour of 0.34 mrad, the resulting Skyrme number obtained from our numerical results is $N^{21}_\mathrm{S}=0.957$, $N^{23}_\mathrm{S}=1.000$, $N^{25}_\mathrm{S}=0.948$.

\begin{figure}
	\centering
	\includegraphics[width=0.48\textwidth]{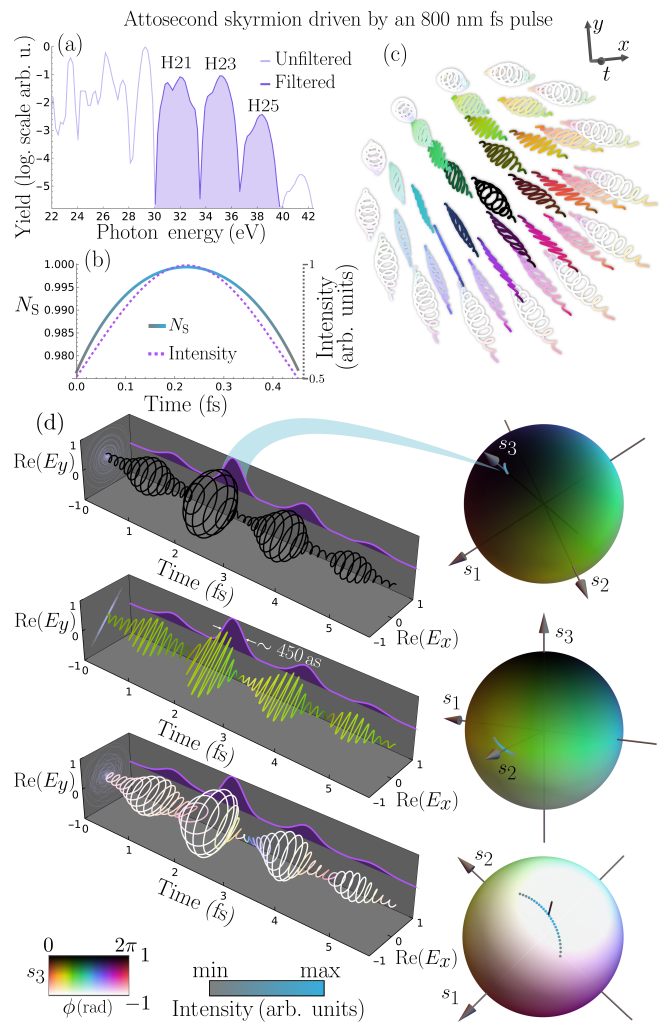}
	\caption{
	Attosecond EUV skyrmion pulse train driven at 0.8 $\mu$m. 
	(a) HHG spectrum, where the purple filled window contains the harmonics $q=21,23,25$ that yield the skyrmion attosecond train.
        (b) Temporal evolution of $N_\mathrm{S}$ inside a closed contour of 0.34~mrad corresponding to the central pulse of the train.
	(c) Local temporal evolution of the real electric field for the central pulse within the train, with the instantaneous polarization state encoded by a color map that maps each point to its location on the Poincaré sphere. 
	(d) Polarization dynamics at representative transverse positions (beam center, intermediate point, and skyrmion edge), with trajectories on the Poincaré sphere within the central pulse; bluish regions indicate the highest pulse intensities.
	}
	\label{fig:figure2}
\end{figure}

We next analyze the resulting attosecond pulses and the temporal stability of their polarization texture. Figure \ref{fig:figure2}(a) shows the spatially-integrated far-field spectrum. Selecting a spectral window containing harmonics $q=21,23,25$ yields a skyrmion attosecond pulse train. Figure~\ref{fig:figure2}(b) shows the temporal evolution of $N_\mathrm{S}$ within a closed contour of 0.34 mrad for the central pulse within the train, reaching $N_\mathrm{S}=1.000$ at the pulse peak, with variations smaller than $\Delta N_\mathrm{S}=0.024$ towards the edges.
The temporal evolution of the local real electric field for a single pulse within the train is shown in Fig.~\ref{fig:figure2}(c). At each point, the field traces in time an approximately constant polarization ellipse. Small instantaneous deviations are encoded by a color scheme that maps the instantaneous polarization state to its point on the Poincaré sphere. 
Such small deviations are further evidenced in the three paradigmatic transverse positions depicted in Fig.~\ref{fig:figure2}(d), which shows the temporal evolution of the attosecond pulse train at positions where the polarization state of the central harmonic ($q=23$) is LCP (top), linear at $45^\circ$ (middle), and RCP (bottom). The central pulse within the train exhibits a duration of $\sim 450$ as (FWHM). The instantaneous polarization trajectories within the FWHM of the central pulse are mapped on the Poincaré sphere (right column). The dots represented on the surface indicate that the polarization state varies slightly along the pulse.

\begin{figure}
	\includegraphics[width=0.48\textwidth]{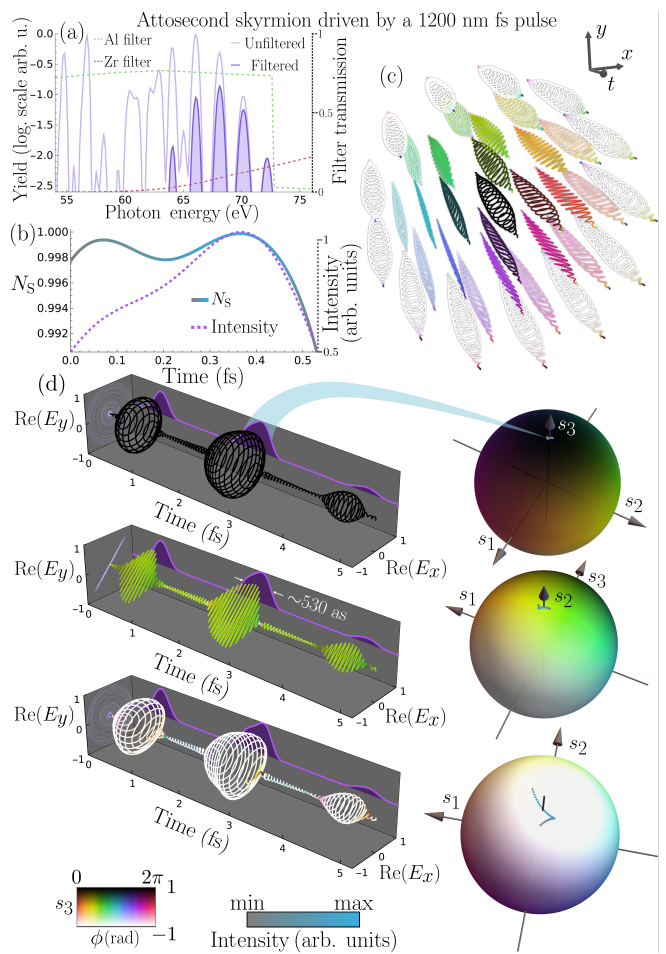}	
	\caption{
	Attosecond EUV skyrmion pulse train driven at 1.2 $\mu$m. 
	(a) HHG spectrum, where the filled purple region shows the resulting harmonic window after using Al and Zr filters that yields the skyrmion attosecond train.
    (b) Temporal evolution of $N_\mathrm{S}$ within the 0.19~mrad contour during the central pulse.
	(c) Local temporal evolution of the real electric field for the central pulse within the train, with the instantaneous polarization state encoded by a color map linking each point to its position on the Poincaré sphere. 
	(d) Polarization dynamics at representative transverse positions (beam center, intermediate point, and skyrmion edge), with trajectories on the Poincaré sphere within the central pulse; bluish regions mark the highest pulse intensities.
	}
	\label{fig:figure3}	
\end{figure}

In order to provide a more realistic experimentally realizable configuration, where the harmonic skyrmions with similar $N_\mathrm{S}$ within the same spatial contour can be naturally filtered, we perform simulations with a driving beam with longer, mid-IR wavelength, of 1.2 $\mu$m, 11.5 fs FWHM and peak intensity of 1.19$\times$10$^{14}$ W/cm$^2$. Increasing the driver wavelength increases the maximum HHG photon energy \cite{Popmintchev2012, Cousin2014, Johnson2018} while reducing the polarization imperfections between consecutive harmonic skyrmions. In Fig.~\ref{fig:figure3}(a) we show the resulting integrated spectrum (purple solid line). The use of $0.2$ $\mu$m Al and $0.3$ $\mu$m Zr foils allows to naturally select a spectral region at 70 eV, covering five harmonic orders (purple filled).

We chose $q=66$ as the central frequency of the resulting window, for which, the resulting driver topological charges are $\ell_{l,r}=-65/132,67/132$, as dictated by Eq.~\ref{eq:input_charges}. Within a closed contour of 0.19 mrad, the Skyrme numbers of the five harmonics within the window, ordered by increasing harmonic energy, are 0.982, 0.995, 1.000, 0.994, and 0.978, showing smaller deviations from the central harmonic than in the 0.8-$\mu$m driver case. The temporal evolution of $N_\mathrm{S}$ within this contour during the central pulse is shown in Fig.~\ref{fig:figure3}(b), yielding $N_\mathrm{S}=1.000$ at the pulse maximum with a maximum deviation of $\Delta N_\mathrm{S}=0.009$.
The resulting skyrmion pulse train is shown in Figs.~\ref{fig:figure3}(c,d). The skyrmion EUV attosecond pulses not only possess higher photon energy, but the polarization state in each spatial position is more regular in time, as evidenced by the corresponding projections on the Poincar\'e sphere.



As a conclusion, we have unveiled a scenario of HHG capable of generating attosecond skyrmion pulses. Despite the harmonic-dependence of the linearly polarized vector driving field, our numerical simulations demonstrate that high-quality attosecond skyrmionic polarization textures emerge when multiple harmonics are combined, an effect particularly favorable using mid-IR drivers. The resulting EUV attosecond skyrmion pulses enable advanced opportunities to control photoelectron distributions \cite{younis2024spin}, to high resolution imaging with EUV structured light \cite{Pancaldi2024, Wang2023, Du2023, Esashi2018}, topological harmonic spectroscopy \cite{GarciaCabrera2024}, the exploration of chiral topological systems \cite{Mayer2024} and magnetic helicoidal dichroism \cite{Fanciulli2022}.

\section*{Acknowledgements}

We acknowledge Rodrigo Mart\'in-Hern\'andez and Charles Durfee for fruitful discussions. This project has received funding from the European Research Council (ERC) under the European Union’s Horizon 2020 research and innovation programme (grant agreement No 851201) and from the Department of Education of the Junta de Castilla y León and FEDER Funds (Escalera de Excelencia CLU-2023-1-02 and grant No. SA108P24). We acknowledge funding from Ministerio de Ciencia e Innovacion (Grant PID2022-142340NB-I00). 

\clearpage

\bibliography{apssamp}	
\bibliographystyle{unsrt}

\end{document}